\documentstyle[pra,aps,array,amsfonts,eqsecnum,12 pt]{revtex}              

\draft
\begin{document}
\bibliographystyle{plain}

\title{Solutions to the Optical Cascading Equations}
\author{S. \textsc{Lafortune} and P. Winternitz}

\address{
 Centre de Recherches Math{\'e}matiques,\\
 Universit{\'e} de Montr{\'e}al,\\
 C. P. 6128, Succ.~Centre-ville,\\
Montr{\'e}al, (QC) H3C 3J7,\\
Canada
}

\author{C.R. Menyuk}

\address{
Departement of Electrical Engineering\\
University of Maryland\\
Baltimore, Maryland\\
21228-5398}

\maketitle

\def\sech{\rm sech}
\def\sec{\rm sec}
\def\cosech{\rm cosech}
\def\cosec{\rm cosec}
\def\cotanh{\rm cotanh}
\def\tanh{\rm tanh}
\def\sn{\rm sn}
\def\ns{\rm ns}
\def\cn{\rm cn}
\def\dn{\rm dn}
\def\tn{\rm tn}
\def\Ima{\rm Im}
\def\Re{\rm Re}

\begin{abstract}
Group theoretical methods are used to study the equations describing 
$\chi ^{(2)} : \chi ^{(2)}$ cascading. The equations are shown not to be 
integrable
by inverse scattering techniques. On the other hand, these equations do share 
some of
the nice properties of soliton equations. Large families of explicit analytical
solutions are obtained in terms of elliptic functions. In special cases these 
periodic
solutions reduce to localized ones, i.e. solitary waves. All previously known 
explicit
solutions are recovered and many new ones are obtained.
\end{abstract}


\newpage

\section{Introduction}

Materials with a significant $\chi^{(3)}$ nonlinearity exhibit solitonlike
beam or
pulse propagation.
These materials can be modeled  by the
nonlinear Schr\"odinger equation and its
variants [1] which are known to be integrable and possess
exact soliton solutions.

In the case of $\chi^{(2)}$ materials, it has been possible to produce
solitary waves through
$\chi^{(2)}$:$\chi^{(2)}$ cascading [2]. Moreover, several particular cases
of solitary wave
 solutions of the system describing this phenomena have been obtained
numerically
and analytically ([3-9]).

The solitons of $\chi^{(3)}$ type are known to belong to families
that include  periodic solutions.
It is essential to know if this is the case for the solitary waves of
the $\chi^{(2)}$ type, mainly because
the system describing them is not integrable. We do not know what
properties of integrable systems
these solitary wave solutions possess.

Moreover, the periodic solutions containing these solitary waves as limiting
cases 
can be important in their own right, specially if they are stable. For
instance they may propagate as a background to solitary wave signals.

 In this paper we find families of periodic solutions expressed in terms of 
elliptic
functions. As special limiting cases we obtain elementary trigonometric 
solutions and
also localized solitary waves (including all those found earlier). We mention 
that not
all of the solitary waves are stable [10-12] and that the stability of periodic
solutions needs a separate study.

In Section 2 we present the $\chi^{(2)}$:$\chi^{(2)}$ cascading equations and 
find
their Lie point symmetry group. We then reduce the system to a set of four 
coupled
real ordinary differential equations for traveling wave solutions. These reduced
equations are further studied in Section 3. We show that in general the 
equations do
not have the Painlev\'e property [13-14] so the cascading equations are not
integrable. We do however show that well behaved analytical solutions exist in 
special cases. Explicitly, elliptic function solutions are presented in 
Section 4 
together with their elementary function limits. Some further special cases are
discussed in Section 5. Conclusions are drawn in the final Section 6.

\section{The Cascading Equations and their Reduction}

Let us write the cascading equations in the normalized form [3]
\begin{equation}
\begin{array}{cc}
\displaystyle{ia_{1t}-\frac{r}{2}a_{1xx}+a_1^*a_2=0,}&\\
\displaystyle{ia_{2t}-\beta a_2-i\delta
a_{2x}-\frac{\alpha}{2}a_{2xx}+a_{1}^2=0,}&
\label{eq:1}
\end{array}
\end{equation}
where $a_1$ and $a_2$ are the (complex) envelopes of the
fundamental and second harmonic wave, respectively. They are functions of two
variables, namely $t$, the normalized distance along the wave guide and $x$, the
normalized transverse coordinate. The constants involved are $r=\pm1$ and 
$\alpha$,
$\beta$, $\delta$ (all reals).

To obtain explicit solutions we use the method of symmetry reduction, i.e. use
Lie point symmetries to reduce the equations (\ref{eq:1}) to ordinary 
differential
equations. We first replace eq.(2.1) by four real equations, putting
\begin{equation}
\label{eq:2}
a_k(x,t)=R_k(x,t)e^{i\phi_k(x,t)}\,,\;k=1,2,
\end{equation}
$$
0\leq R_k<\infty,\;\;0\leq\phi_k < 2\pi,
$$
and apply a standard algorithm to find their symmetry group [15,16,17].
In the generic case, when all constants in the equation are arbitrary, the 
symmetry
group of the equations is just three-dimensional, consisting of translations
in $x$ and $t$ and corelated shifts of the phase $\phi_1\rightarrow
\phi_1+\lambda$,  $\phi_2\rightarrow
\phi_2+2\lambda$. The Lie algebra of the symmetry group has a basis 
consisting of
the operators
\begin{equation}
P_1=\partial_x,\;\;
P_0=\partial_t,\;\;
W=\partial_{\phi_1}+2\partial_{\phi_2}.
\end{equation}
We shall consider solutions invariant under the subgroup generated by
\begin{equation}
X=vP_1+\omega P_0+(\kappa_1\omega-\omega_1v)W,
\end{equation} 
where $v$, $\omega$, $\kappa_1$ and $\omega_1$ are real constants.

 The invariant solutions then have the
 form
\begin{equation}
\label{eq:5}
\begin{array}{l}
\displaystyle{a_1(\xi)=R_1(\xi)e^{i\phi_1(\xi)}e^{i(\kappa_1t-\omega_1x)},}\\
\\
\displaystyle{a_2(\xi)=R_2(\xi)e^{i\phi_2(\xi)}e^{2i(\kappa_1t-\omega_1x)},}
\end{array}
\end{equation}
$$\xi=\omega x-vt.$$
The four real functions $R_1$, $R_2$, $\phi_1$, $\phi_2$ satisfy the 
following
system of coupled ordinary differential equations
\begin{eqnarray}
\label{eq:8}
&-\frac{r}{2}\omega^2(2R_1'\phi_1'+R_1\phi_1'')+
(r\omega\omega_1-v)R_1'+R_1R_2\sin{\phi}=0,\\
\label{eq:9}
&-\frac{\alpha}{2}\omega^2(2R_2'\phi_2'+R_2\phi_2'')+
(2\alpha \omega \omega_1-v-\delta\omega)R_2'-R_1^2\sin{\phi}=0,\\
\label{eq:10}
&-\frac{r}{2}\omega^2(R_1''-R_1{\phi_1'}^2)+(v-r\omega\omega_1)R_1\phi_1'+
(\frac{r}{2}\omega_1^2-\kappa_1)R_1\nonumber \\
&+R_1R_2\cos{\phi}=0,\\
\label{eq:11}
&-\frac{\alpha}{2}\omega^2(R_2''-R_2{\phi_2'}^2)-
(2\alpha\omega\omega_1-v-\delta\omega)R_2\phi_2' \nonumber\\
&+(2\alpha\omega_1^2-2\kappa_1-\beta-2\delta\omega_1)R_2 
+R_1^2\cos{\phi}=0,
\end{eqnarray}
$$\phi\equiv\phi_2-2\phi_1.$$ 

It is the system of equations (2.6)-(2.9) that we wish to solve.

\section{Analysis of the Reduced Equations}

\subsection{Phase locked solutions}

The system (2.6),...,(2.9) is in general quite difficult to decouple and solve.
It is greatly simplified if we impose a supplementary restriction on the
phases, namely
\begin{equation}
\phi=k\pi,\;\;k\in {\mathbb Z},\;\; {\mbox i.e.}\;\;\sin{\phi}={0},\;\;
\cos{\phi}=\epsilon,
\;\;\epsilon^{2}={1}.
\end{equation}
We shall call such solutions ``phase locked solutions''. Let us first assume
\begin{equation}
r\alpha \omega \neq 0
\end{equation}
and simplify notations, putting
\begin{equation}
\begin{array}{ll}
\displaystyle{A=\frac{r\omega\omega_1-v}{r\omega^2},}&
\displaystyle{B=\frac{2\alpha\omega\omega_1-v-\delta\omega}{\alpha\omega^2},}\\
\\
\displaystyle{C=2\frac{2\alpha\omega_1^2-2\kappa_1-2\delta\omega_1-\beta}
{\alpha\omega^2},}&
\displaystyle{D=\frac{r\omega_1^2-2\kappa_1}{r\omega^2},}
\end{array}
\end{equation}
$$
M_1=-\frac{2N_1}{r\omega^2},\;\;M_2=-\frac{2N_2}{\alpha\omega^2}.
$$
The equations (2.6),...,(2.9) can be rewritten as
\begin{eqnarray}
&\phi_1'=A+\frac{M_1}{R_1^2},\\
&\phi_2'=B+\frac{M_2}{R_2^2},\\
&R_1''-R_1{\phi_1'}^2+2AR_1\phi_1'-DR_1-\frac{2\epsilon}{r\omega^2}R_1R_2=0,\\
&R_2''-R_2{\phi_2'}^2+2BR_2\phi_2'-CR_2-\frac{2\epsilon}{\alpha\omega^2}R_1^2=0.
\end{eqnarray}

\subsection{Case $M_1M_2\neq 0$}

The phase locking condition (3.1) imposes a relation between $R_1$ and $R_2$, 
namely
\begin{equation}
B+\frac{M_2}{R_2^2}=2A+\frac{2M_1}{R_1^2},
\end{equation}
so that the system (3.4),...,(3.8) is overdetermined. Expressing $R_1$, $\phi_1'$
and $\phi_2'$ in terms of $R_2$ and substituing into (3.6) and (3.8), we 
obtain two
second order ordinary differential equations for $R_2$. These turn out to be
compatible only for $R_1$ and $R_2$ constant. This case will be considered 
separately
in Section 5 below.

\subsection{Case $M_1M_2=0$}

Again $R_1$ and $R_2$ are constant, unless we have
\begin{equation}
M_1=M_2=0.
\end{equation}
Let us investigate case (3.9). We have
\begin{equation}
\phi_1'=A,\;\;\phi_2'=B=2A.
\end{equation}
We put
\begin{equation}
A_0=A^2-D,\;\;B_0=-\frac{2\epsilon}{r\omega^2},\;\;C_0=B^2-C,\;\;
D_0=-\frac{2\epsilon}{\alpha \omega^2},
\end{equation}
and obtain a system of two ordinary differential equations
\begin{equation}
\begin{array}{l}
R_1''+A_0R_1+B_0R_1R_2=0,\\ \\
R_2''+C_0R_2+D_0R_1^2=0.
\end{array}
\end{equation}
This system is not overdetermined since the constraint (3.8) is now simply the
condition (3.10) on the constants, i.e.
\begin{equation}
2\alpha v -rv-\delta r \omega=0.
\end{equation}
 Eliminating $R_2$ from eq.(3.12), we obtain a fourth order equation for $R_1$, 
namely
\begin{equation}
\left(\frac{R_1''}{R_1}\right)''+C_0\frac{R_1''}{R_1}-B_0D_0R_1^2+A_0C_0=0.
\label{eq:18}
\end{equation}
If the original equations (2.1) are integrable then eq.(3.14) should have the
Painlev\'e property, i.e. have no movable singularities other than poles [13,14].
 An
algorithmic test exists [13,18] establishing certain properties of an equation, 
necessary
for it to have the Painlev\'e property.

  Thus, the general solution of eq.(3.14) must allow an expansion in the 
neighbourhood
of any singular point of the form
\begin{equation}
R_1=\sum_{k=0}^{\infty} a_k(\xi-\xi_0)^{k+p}
\end{equation}
with $p$ a negative integer, $a_0\neq 0$ and three of the coefficients $a_k$
arbitrary. Then $R_1$ has a good chance of representing the general solution of 
eq.(3.14), depending on four
arbitrary constants (one of them being $\xi_0$, the position of the pole). The 
values
of $k$ for which $a_k$ are arbitrary (i.e. are not determined by a recursion
relation), are called ``resonance'' values.

Substituting the expansion (3.15) into eq.(3.14), we find $p=-2$, 
$a_0^2=36/(B_0D_0)$ and
the resonance values 
\begin{equation}
r=-1,\; 6,\; (5\pm i\sqrt{23})/2.
\end{equation}
Thus, we have only one nonnegative integer, namely $r=6$, rather
than the three ones needed. An 
ana\-ly\-sis of the obtained
recursion relations shows that $a_0$, ..., $a_5$ are fully determined, $a_6$ 
is indeed
free and can be chosen arbitrarily. Then $a_7$ and all the higher terms are fully
determined in terms of $\xi_0$ and $a_6$ (and of course $A_0$, $B_0$, $C_0$ 
and $D_0$).

Thus, eq.(3.14) does not have the Painlev\'e property and the cascading 
equations (2.1)
are not integrable.

The Painlev\'e analysis does however indicates that families of ``well
behaved'' 
solutions should exist (i.e. single valued in the neighbourhood of their movable 
singularities), depending on one or two free parameters rather than on four.
We shall find such solutions below.

An alternative procedure is to solve eq.(3.12), again under the condition 
(3.13), for
$R_2$. We obtain the ordinary differential equation
\begin{eqnarray}
2(R_2''+C_0R_2)(R_2''''+C_0R_2'')-(R_2'''+C_0R_2')^2
\nonumber \\
+4(A_0+B_0R_2)(R_2''+C_0R_2)^2=0.
\end{eqnarray}
A Painlev\'e analysis of eq.(3.17) leads to the same conclusion as that of 
eq.(3.14).

\subsection{Introduction of the Elliptic Function Equation}

Let us look for solutions $R_2(\xi)$ satisfying eq.(3.17) and also the elliptic
function equation
\begin{equation}
{R_2'}^2=\gamma_4R_2^4+\gamma_3R_2^3+\gamma_2R_2^2+\gamma_1R_2+\gamma_0.
\end{equation}
The compatibility of eq.(3.18) and (3.17) implies 
\begin{equation}
\gamma_4=0
\end{equation}
and imposes six relations between the constants in (3.18) and (3.17). 
These allow for
the following solutions.
\begin{equation}
\hspace{-5cm}{\rm 1.}\;\; \gamma_3\neq 0,\;\;C_0(C_0-A_0)\neq 0.
\end{equation}
The constants $\gamma_\mu$ in eq.(3.18) are completely specified:
\begin{equation}
\begin{array}{c}
\displaystyle{\gamma_3=-\frac{2B_0}{3},\;\;\gamma_2=C_0-2A_0,\;\;
\gamma_1=\frac{(C_0+2A_0)(C_0-A_0)}{B_0},}\\
\displaystyle{\gamma_0=\frac{(C_0+2A_0)^2(C_0-A_0)}{6B_0^2}.}
\end{array}
\end{equation}
The other amplitude, $R_1(\xi)$ is given directly by the expression
\begin{equation}
R_1^2=\frac{1}{D_0}\Big{(}B_0R_2^2+2(A_0-C_0)R_2+\frac{(A_0-C_0)(2A_0+C_0)}{2B_0}
\Big{)}
\end{equation}
and we must require $R_1^2$ to satisfy
\begin{equation}
R_1^2\geq 0
\end{equation}
in the entire range of values of $R_2$, a condition to be analysed below.

Notice that eq.(3.18) will have solutions depending on just one parameter, an
integration constant, since the coefficients $\gamma_\mu$ are fixed in terms of 
$A_0$,
$B_0$, $C_0$ and $D_0$. The only constraints on the constants in the original
equations (2.1) $r$, $\alpha$, $\beta$ and $\delta$ and those introduced
in the reduction procedure (2.5) namely $v$, $\omega$, $\kappa_1$ and
$\omega_1$ are (3.2), (3.13) and also (3.23).
\begin{equation}
\hspace{-7cm}{\rm 2.}\;\; \gamma_3\neq 0,\;\;C_0=0.
\end{equation}
In this case we obtain 
\begin{equation}
\gamma_3=-\frac{2B_0}{3},\;\;\gamma_2=-2A_0,\;\;\gamma_1=-\frac{2A_0^2}{B_0}
\end{equation}
and $\gamma_0$ is arbitrary. Moreover, we have
\begin{equation}
R_1=\sqrt{\frac{B_0}{D_0}}\Big{(}R_2+\frac{A_0}{B_0}\Big{)}.
\end{equation}
Thus, we obtain a two parameter family of solutions, the parameters being 
$\gamma_0$
and a constant arising in the integration of eq.(3.18). The constraints on the
coefficients are (3.2), (3.13) and $C_0=0$, i.e.
\begin{equation}
2\alpha v^2+\omega (2\kappa_1 \omega+\beta \omega-2\omega_1 v)=0
\end{equation}
(we have $r^2=1$).
\begin{equation}
\hspace{-7cm}{\rm 3.}\;\; \gamma_3\neq 0,\;\;C_0=A_0.
\end{equation}
We then have
\begin{equation}
\gamma_3=-\frac{2B_0}{3},\;\;\gamma_2=-A_0,\;\;\gamma_1=0
\end{equation}
and $\gamma_0$ arbitrary. Again, (3.18) provides a two parameter family of 
solutions
and the constraint $C_0=A_0$ is
\begin{equation}
3\alpha v^2+2r\omega(\alpha-2r)(\omega_1 v-\kappa_1 \omega)+2\beta\omega^2=0
\end{equation}
and we have
\begin{equation}
R_1=\sqrt{\frac{B_0}{D_0}}R_2.
\end{equation}
\begin{equation}
\hspace{-8cm}{\rm 4.}\;\; \gamma_3=0.
\end{equation}
In this case eq.(3.18) reduces to
\begin{equation}
{R_2'}^2=-C_0R_2+\gamma_0.
\end{equation}
The solutions are
\begin{equation}
R_1=0
\end{equation}
\begin{equation}
R_2=
\left\{ \normalsize
\begin{array}{ll}
\displaystyle{-\frac{C_0}{4}(\xi-\xi_0)^2+\mu}&
\displaystyle{C_0\neq 0}\\
\displaystyle{\mu (\xi-\xi_0)}&
\displaystyle{C_0=0.}
\end{array}
\right.
\end{equation}

\subsection{Phase Locked Solutions for $\alpha=0$}

We return to eq.(2.6),...,(2.9) for $\alpha=0$, $\sin\phi=0$, 
$\cos\phi=\epsilon$. From
eq.(2.7) we see that $R_2\neq {\rm const}$ implies
\begin{equation}
v+\delta \omega=0.
\end{equation}
Eq.(2.6) can be integrated to give
\begin{equation}
\phi_1'=\frac{2N}{r\omega^2}\frac{1}{R_1^2}+\frac{r\omega\omega_1 -v}{2},
\end{equation}
where $N$ is an integration constant.

In order to have $R_1\neq 0$ we impose
\begin{equation}
2\kappa_1+\beta+2\delta\omega_1\neq 0
\end{equation}
and obtain from eq.(2.9) that we have
\begin{equation}
R_1=[\epsilon(2\kappa_1+\beta+2\delta\omega_1)R_2]^{1/2}.
\end{equation}
Finally, eq.(2.8) implies that $R_2$ satisfies the elliptic function equation
\begin{equation}
(R_1')^2=-2B_0R_2^3-4A_0R_2^2+SR_2-\frac{16N^2}{\omega^4(2\kappa_1+\beta+2\delta
\omega_1)^2}.
\end{equation}
Since both $N$ and $S$ are arbitrary integration constants, eq.(3.40) yields a 3
parameter family of solutions.

\section{Solutions in terms of Elliptic Functions and their Limiting Cases}

\subsection{General Comments on the Elliptic Function Equation}

In Section 3 we have obtained three equations of the type
\begin{equation}
{R'}^2=\beta_3R^3+\beta_2R^2+\beta_1R+\beta_0,
\label{eq:37}
\end{equation}
where $\beta_i$ are real constants and  $\beta_3\neq 0$. Putting
\begin{equation}
f=\beta_3R
\end{equation}
we obtain the equation
\begin{equation}
{f'}^2=f^3+\beta_2f^2+\beta_1\beta_3f+\beta_0\beta_3^2.
\label{eq:38}
\end{equation}
We introduce the roots $f_1$, $f_2$ and $f_3$ of the polynomial on the
right hand side of eq.(4.3) and rewrite this equation as
\begin{equation}
{f'}^2=(f-f_1)(f-f_2)(f-f_3).
\label{eq:39}
\end{equation}
The solutions of this equation can be expressed in terms of Jacobi elliptic
functions [19] if the three roots are distinct. The case of multiple roots 
leads to
solutions in terms of elementary functions.

Let us first consider the case of three real roots and order them to satisfy 
$f_1\leq f_2 \leq f_3$. We are only interested in real solutions.
$$
\hspace{-6cm} {\rm 1.}\;\; f_1\leq f \leq f_2 < f_3 
$$
We obtain a finite periodic solution
\begin{equation}
f=(f_2-f_1)\sn^2{(u,k)}+f_1,
\label{eq:43}
\end{equation}
$$
k^2=\frac{f_2-f_1}{f_3-f_1},\;\;
u=\frac{\sqrt{f_3-f_1}}{2}(\xi-\xi_0),
$$
where $\xi_0$ is a real integration constant.
$$
\hspace{-6cm} {\rm 2.}\;\; f_1\leq f\leq f_2 = f_3 
$$
A special (limiting) case of solution (4.5) is obtained for $f_2=f_3$, namely the
solitary wave solution
\begin{equation}
f=(f_2-f_1)\tanh^2{(u)}+f_1,
\label{eq:44}
\end{equation}
with $u$ as in equation (4.5).
$$
\hspace{-6cm} {\rm 3.}\;\; f_1 < f_2 < f_3 \leq f
$$
We obtain a singular periodic solution
\begin{equation}
f=(f_3-f_1)\frac{1}{\sn^2{(u,k)}}+f_1,
\label{eq:45}
\end{equation}
with $u$ and $k$ as in eq.(4.5).
$$
\hspace{-6cm} {\rm 4.}\;\; f_1 = f_2 < f_3 \leq f 
$$
For $f_1=f_2$ solution (4.7) reduces to an elementary periodic singular 
solution namely
\begin{equation}
f=(f_3-f_1)\frac{1}{\sin^2{(u)}}+f_1.
\label{eq:46}
\end{equation}
$$
\hspace{-6cm} {\rm 5.}\;\; f_1 < f_2 = f_3 \leq f 
$$
For $f_2=f_3$ solution (4.7) reduces to a ``singular solitary wave'', namely
\begin{equation}
f=(f_3-f_1)\cotanh^2{(u)}+f_1.
\label{eq:47}
\end{equation}
$$
\hspace{-6cm} {\rm 6.}\;\; f_1 = f_2 = f_3 \leq f 
$$
A triple root corresponds to a ``singular algebraic solitary wave''
\begin{equation}
f=\frac{4}{(\xi-\xi_0)^{2}}+f_1.
\label{eq:48}
\end{equation}

The case of one real and two mutually complex conjugated roots leads to singular
periodic solutions.
$$
\hspace{-3cm} {\rm 7.}\;\; f_1 = a+ib,\;\;f_2 =a-ib,\;\; f_3,a,b \in 
{\mathbb R},\;b>0 
$$
The solution is
\begin{equation}
f=\frac{f_3+P+(f_3-P)\cn{(u,k)}}{1+\cn{(u,k)}}, 
\label{eq:52}
\end{equation}
\begin{equation}
u=\sqrt{P}(\xi-\xi_0),\;\;k^2=\frac{P+a-f_3}{2P},\;\;
P^2=(a-f_3)^2+b^2.
\end{equation}
In the limit $b\rightarrow 0$ solution (4.11) reduces to the solitary wave (4.6).

\subsection{Explicit Solutions of the Cascading Equations for $\alpha\neq 0$}
$$
\hspace{-7cm} {\rm 1.\; The\;\; case\;\; (3.20)} 
$$
We have
\begin{equation}
\begin{array}{c}
\displaystyle{\beta_2=C_0-2A_0,\;\;
\beta_1\beta_3=-\frac{2}{3}(C_0+2A_0)(C_0-A_0)}\\
\displaystyle{\beta_0\beta_3^2=\frac{2}{27}(C_0+2A_0)^2(C_0-A_0),
\;\;C_0\neq0,\;\;C_0\neq A_0}
\end{array}
\end{equation}
in eq.(4.3). Using eq.(3.22) we have
\begin{equation}
R_2=-\frac{3}{2 B_0}f,\;\;R_1^2=\frac{9}{4B_0D_0}[f^2-\frac{4}{3}(A_0-C_0)f
+\frac{2}{9}(A_0-C_0)(2A_0+C_0)].
\end{equation}
A globally defined solution exists if $R_1^2$ is positive for all values of $f$. 
We note that the roots of the polynomial defining $R_1^2$ are
\begin{equation}
f_{A,B}=\frac{1}{3}[2(A_0-C_0)\pm\sqrt{6C_0(C_0-A_0)}].
\end{equation}
Hence, for $C_0(C_0-A_0)<0$ the function $R_1^2$ is sign-definite and can 
be chosen
to be positive definite. For $C_0(C_0-A_0)>0$ a more careful analysis is 
required. Thus solution
(4.5) provides a global solution if the roots satisfy one of the following 
relations:
\begin{equation}
\begin{array}{c}
\displaystyle{
f_A\leq f_1\leq f\leq f_2\leq f_B,\;\;f_A\leq f_B\leq f_1\leq f\leq f_2,}\\
\displaystyle{f_1\leq f\leq f_2\leq f_A\leq f_B.}
\end{array}
\end{equation}
Similarly, case (4.7) and (4.11) both require
\begin{equation}
f_A\leq f_B\leq f_3\leq f.
\end{equation}

Multiple roots occur in the two excluded cases $C_0=0$ and $C_0=A_0$, but also 
in the allowed case
\begin{equation}
C_0=-2A_0
\end{equation}
For $A_0<0$ we obtain from eq.(4.6)
\begin{equation}
\begin{array}{c}
\displaystyle
{R_2=\frac{6\left|A_0\right|}{B_0}\frac{1}
{\cosh^2{(
\sqrt{\left|A_0\right|}
(\xi-\xi_0))}}}
\\ \\
\displaystyle{
R_1=\frac{6\left|A_0\right|}{\sqrt{-B_0D_0}}
\frac{\sinh{(\sqrt{\left|A_0\right|}(\xi-\xi_0))}}{
\cosh^2{(\sqrt{\left|A_0\right|}(\xi-\xi_0))}}}
\end{array}
\end{equation}
so that we must require $B_0D_0<0$.

Similarly, for $A_0<0$ eq.(4.9) provides the solution
\begin{equation}
\begin{array}{c}
\displaystyle
{R_2=-\frac{6\left|A_0\right|}{B_0}\frac{1}
{\sinh^2{(
\sqrt{\left|A_0\right|}
(\xi-\xi_0))}}}
\\ \\
\displaystyle{
R_1=\frac{6\left|A_0\right|}{\sqrt{B_0D_0}}
\frac{\cosh{(\sqrt{\left|A_0\right|}(\xi-\xi_0))}}{
\sinh^2{(\sqrt{\left|A_0\right|}(\xi-\xi_0))}}}
\end{array}
\end{equation}
with the requirement $B_0D_0>0$.

For $A_0>0$ it is solution (4.8) that is relevant and yields:
\begin{equation}
\begin{array}{c}
\displaystyle
{R_2=-\frac{6A_0}{B_0}\frac{1}
{\sin^2{(
\sqrt{A_0}
(\xi-\xi_0))}}}
\\ \\
\displaystyle{
R_1=\frac{6A_0}{\sqrt{B_0D_0}}
\frac{\cos{(\sqrt{A_0}(\xi-\xi_0))}}{
\sin^2{(\sqrt{A_0}(\xi-\xi_0))}},\;\;B_0D_0>0}.
\end{array}
\end{equation}
$$
\hspace{-7cm}{\rm 2.\;The\;\;case\;\;(3.25):}\;\;C_0=0 
$$
Eq.(4.3) is
\begin{equation}
{f'}^2=f^3-2A_0f^2+\frac{4}{3}A_0^2f+\gamma_0\frac{4B_0^2}{9}
\end{equation}
with $\gamma_0$ arbitrary and
\begin{equation}
R_2=-\frac{3}{2B_0}f,\;\;R_1=\frac{1}{2\sqrt{B_0D_0}}(-3f+2A_0),\;\;B_0D_0>0.
\end{equation}
A multiple root, namely a triple one, occurs in one case only
\begin{equation}
{f'}^2=(f-\frac{2}{3}A_0)^3,\;\;\gamma_0=-\frac{2A_0^3}{3B_0^2}
\end{equation}
and we have
\begin{equation}
\begin{array}{c}
\displaystyle{R_2=-\frac{1}{B_0}\big(\frac{6}{(\xi-\xi_0)^2}+A_0\big),}\\ \\
\displaystyle{R_1=\frac{-6}{\sqrt{B_0D_0}}\frac{1}{(\xi-\xi_0)^2}.}
\end{array}
\end{equation}

In all other cases two of the roots $f_i$ are complex and $f$ is given 
in eq.(4.11),(4.12).
$$
\hspace{-7cm}{\rm 3.\;The\;\;case\;\;(3.28):}\;\;C_0=A_0 
$$
Eq.(4.3) in this case is
\begin{equation}
{f'}^2=f^3-A_0f^2+\gamma_0\frac{4B_0^2}{9}
\end{equation}
where $\gamma_0$ is arbitrary and we have
\begin{equation}
R_2=-\frac{3}{2B_0}f,\;\;R_1=-\frac{3}{2\sqrt{B_0D_0}}f,\;\;
B_0D_0>0.
\end{equation}
The discriminant of the cubic equation ${f'}^2=0$ is
\begin{equation}
D=\frac{16}{9}(A_0^3-3B_0^2\gamma_0)B_0^2\gamma_0.
\end{equation} 
For $D<0$ two roots are complex, for $D>0$ they are all real and distinct, 
for $D=0$ we have a double or triple root. The absence of a first degree term
in (4.26) implies that for $f_i$ real the three signs of $f_i$ cannot be all 
the same. Moreover, if a root is equal to $f_i=0$, it must be double.

The elementary solutions in this case are:
\begin{enumerate}
\item  $\gamma_0=\frac{A_0^3}{3B_0^2}.$ The finite solitary wave, the 
singular solitary wave, the singular periodic solution and the singular 
algebraic solutions 
in this case are:
\begin{equation}
R_2=\sqrt{\frac{D_0}{B_0}}
R_1=
-\frac{A_0}{2B_0}
[3\tanh^2{\big(\frac{\sqrt{A_0}}{2}
(\xi-\xi_0)\big)}-1],\;\;A_0>0
\end{equation}
\begin{equation}
R_2=\sqrt{
\frac{D_0}{B_0}}R_1=-\frac{A_0}{2B_0}[3\cotanh^2{\big(\frac{\sqrt{A_0}}{2}
(\xi-\xi_0)\big)}-1],\;\;A_0>0
\end{equation}
\begin{equation}
R_2=\sqrt{\frac{D_0}{B_0}}R_1=-\frac{\left|A_0\right|}{2B_0}
\left[\frac{3}{\sin^2{\big(\frac{\sqrt{\left|A_0\right|}}{2}
(\xi-\xi_0)\big)}}-2\right],\;\;A_0<0
\end{equation}
\begin{equation}
R_2=\sqrt{\frac{D_0}{B_0}}R_1=-\frac{6}{B_0(\xi-\xi_0)^2},\;\;A_0=0
\end{equation}
respectively.
\item  $\gamma_0=0.$ The four different real solutions 
in this case are:
\begin{eqnarray}
R_2=\frac{3}{2B_0}\left|A_0\right|\sech^2{(\frac{\sqrt{\left|A_0\right|}}{2}
(\xi-\xi_0))},\;\;A_0<0\\
R_2=-\frac{3}{2B_0}\left|A_0\right|\frac{1}{
\sinh^2{(\frac{\sqrt{\left|A_0\right|}}{2}
(\xi-\xi_0))}},\;\; A_0<0\\
R_2=-\frac{3A_0}{2B_0}\frac{1}{\sin^2{(\frac{\sqrt{A_0}}{2}
(\xi-\xi_0))}},\;\; A_0>0
\end{eqnarray}
and for $A_0=0$ we reobtain solution (4.31).
\end{enumerate}

\subsection{Solutions for $\alpha=0$}
Eq.(4.3) (obtained from eq.(3.40)) in this case is
\begin{equation}
{f'}^2=f^3-4A_0f^2-2B_0Sf-4M^2B_0^2
\end{equation}
where $S$ and $M$ are arbitrary. All possibilities for the
three roots $f_i$ occur, but we have the restriction
\begin{equation}
f_1f_2f_3\geq 0.
\end{equation}
We have
\begin{equation}
R_2=-\frac{1}{2B_0}f,\;\;R_1=\sqrt{-\epsilon\frac{
(2\kappa_1+\beta+2\delta\omega_1)}{2B_0}f}.
\end{equation}
For $R_1$ to be globally defined we need $f$ to be sign definite. 
This imposes the 
following restrictions on the roots for each of the solutions of section (4.1). 

{\noindent}
Solutions
(4.5) and (4.6):
\begin{equation}
f_1\leq f\leq f_2\leq 0\leq f_3\;\;{\rm or}\;\;0\leq f_1\leq f\leq f_2 \leq f_3.
\end{equation}
Solutions (4.7), (4.8) and (4.9):
\begin{equation}
0\leq f_1\leq f_2\leq f_3 \leq 
f\;\;{\rm or}\;\;f_1\leq f_2\leq 0 \leq f_3 \leq f.
\end{equation}
Solution (4.10):
\begin{equation}
0 \leq f_1 = f_2=f_3\leq f. 
\end{equation}
Solution (4.11):
\begin{equation}
0\leq f_3 \leq f.
\end{equation}

\section{Other Explicit Solutions}
In Section 4 we presented solutions of eq.(2.1) satisfying $a_1a_2\neq 0$ and 
such that 
$\left|a_1\right|$ and $\left|a_2\right|$ are not constant. Let us now discuss 
these
previously rejected solutions.
$$
\hspace{-6cm}{\rm 1.\;\;Solutions\;\;with}\;\;a_1=0 
$$

For $\alpha\neq 0$ we put
\begin{equation}
a_2(x,t)=\omega(x,t)
\exp{i\big[-(\beta+\frac{\delta^2}{2\alpha})t-\frac{\delta}{\alpha}x\big]}.   
\end{equation}
The eq.(2.1) for $a_2(x,t)$, $a_1=0$ reduces to the linear Schr\"odinger equation
\begin{equation}
i\omega_t-\frac{\alpha}{2}\omega_{xx}=0.
\end{equation}

For $\alpha=0$ the solution of eq.(2.1) is
\begin{equation}
a_2(x,t)=e^{-i\beta t}\omega(\xi),\;\;\xi=x+\delta t
\end{equation}
where $\omega(\xi)$ is an arbitrary (complex) function.
$$
\hspace{-4 cm}{\rm 2.\;\;Solutions\;\;with\;\;}R_1{\;\;and\;\;}R_2{\;\;constant}
$$

We require that $\phi_1$ and $\phi_2$ should not be constant, otherwise we 
obtain $a_1=a_2=0$.
Eq.(2.6),\dots,(2.9) for $R_1$ and $R_2$ constant imply that we must have 
$\sin{\phi}=0$,
i.e. the solutions will be phase locked.

Explicitly, for $\alpha\neq 0$ we have
\begin{equation}
\begin{array}{c}
\displaystyle{R_2=-\frac{r\omega^2}{2\epsilon}(\gamma_1^2-2A\gamma_1+D)}\\ \\
\displaystyle{R_1^2=\frac{\alpha r \omega^4}{4}(4\gamma_1^2-2B\gamma_1+C)
(\gamma_1^2-2A\gamma_1+D)}\\ \\
\displaystyle{\phi_1=\gamma_1\xi+\gamma_2}\\ \\
\displaystyle{\phi_2=2\gamma_1\xi+2\gamma_2+k\pi,\;\;\epsilon=(-1)^k}
\end{array}
\end{equation}
where $\gamma_1$ and $\gamma_2$ are arbitrary constants.

For $\alpha=0$ we have $\phi_1$ and $\phi_2$ as in (5.4)
\begin{equation}
\begin{array}{c}
\displaystyle{R_2=-\epsilon\big[\frac{
r\omega^2}{2}\gamma_1^2+(v-r\omega\omega_1)\gamma_1+
\frac{r\omega_1^2-2\kappa_1}{2}\big]}\\ \\
\displaystyle{
R_1^2=-\epsilon[(v+\delta \omega)\gamma_1-(2\kappa_1+\beta+\delta\omega_1)]R_2}
\end{array}
\end{equation}

\section{Conclusion}

The symmetries used in this article are only those that exist in the generic 
case of the cascading equations (2.1), i.e. for all values of the constants 
$r$, $\alpha$, $\beta$ and $\delta$. This ``generic'' symmetry algebra is 
summed up 
in eq.(2.3), representing space and time translations and a shift in the phases
of the fundamental and second harmonic waves.

The reduction to the system of ordinary differential equations (2.6)...(2.9) was 
achieved by requiring that solutions be invariant under the one-dimensional 
subgroup of the symmetry group, corresponding to the Lie algebra element (2.4).
In order to decouple these equations we had to impose either $a_1=0$ (no 
fundamental harmonic), or the supplementary condition (3.1) on the phases (phase
locked solutions). The existence of the symmetry $W$ in eq.(2.3) guarantees that
if the phases locking condition (3.1) is imposed on the initial conditions, it
will survive for all times.

Thus, by construction, all the explicit solutions obtained in this article are 
phase locked travelling waves. Let us discuss some of their features.

Solution (4.14), for certain values of the parameters $r$, $\alpha$, 
$\beta$, $\delta$, 
characterizing the material involved, and of the constants $\omega$, 
$v$, $\omega_1$ and
$\kappa_1$, characterizing initial conditions, can be periodic 
finite waves (see eq.(4.5)).
The second harmonic $R_2$ then oscillates in the interval ($-\frac{3}{2B_0}f_1$, 
$-\frac{3}{2B_0}f_2$) and the fundamental wave also oscillates between 
finite limits.

For other conditions (see eq.(4.18)) the elliptic function solutions reduce to 
solitary waves (4.19) with the second harmonic going through a zero when the 
fundamental one
reaches its maximum value. Asymptotically both waves (4.19) tend to zero with 
the second harmonic
vanishing at a faster rate.

Many of the obtained solutions are singular, either at some specific point 
$\xi=\xi_0$ as in eq.(4.20), or periodically, as in (4.7), or (4.11).

For conditions leading to eq.(3.25) the second harmonic differs from the
fundamental one just by a proportionality factor and a constant shift of the
amplitude (see eq.(3.26)). Similarly, for conditions (3.28), the two waves
are simply proportional.

Singular solutions coexist with the finite ones for all the values of the
coefficients in eq.(2.1). Their physical meaning needs a separate
investigation: they may be an indication that higher harmonics or dissipative
effects were unjustifiably ignored in the derivation of these equations. These
effects would tend to smoothen out the singularities and possibly turn them
into finite resonance phenomena.

The existence of families of elliptic function solutions can be viewed as a
manifestation of ``partial integrability''. We have shown that the studied
equations do not, in general have the Painlev\'e property. For special values
of the constants involved we do get solutions that do have this property:
they have no movable singularities other than poles.

Finally let us mention that for particular values of the constants in
eq.(2.1) the symmetry group may be larger. For instance, for $\beta=\delta=0$
the equations are invariant under dilatations generated by:
$$
D=x \partial_x+2 t \partial_t-2(R_1 \partial_{R_1}+R_2\partial_{R_2}).
$$
This raises the posibility of obtaining self-similar solutions of the form
$$
\begin{array}{l}
\displaystyle{
a_1(\xi)=\frac{1}{t}F_1(\xi)e^{i\phi_1(\xi)}e^{\frac{i}{2}\ln t},}\\
\\
\displaystyle{
a_2(\xi)=\frac{1}{t}F_2(\xi)e^{i\phi_2(\xi)}e^{i\ln t},}
\end{array}
$$
$$
\xi=\frac{x}{\sqrt{t}}.
$$
Since self-similar solutions are particularly important (and stable) in
optical systems with memory [20,21,22], this situation may be will worth
exploring.

\begin{center}
\large{\bf ACKNOWLEDGEMENTS}
\end{center}
\noindent The authors
 thank D.Levi, Z.Thomova and Yu.S.Kivshar for helpful discussions. S. L. 
acknowledges a scholarship from NSERC
(National Science and
Engineering Research Council of Canada) for his Ph.D studies. The research of 
P.W. was partly supported by research
grants from NSERC of Canada and FCAR du Qu\'ebec.


\begin{thebibliography}{99}


\bibitem{1} See e.g., G.P. Agrawal, {\em Nonlinear Fiber Optics, Academic,
San Diego (1989)}, Chap. 2.

\bibitem{2} W.E.Torruellas, Z.Wang, D.J.Hagan, E.W.Van Strykland, G.I.Stegeman,
L.Torner and C.R.Menyuk, {\em Phys.Rev.E {\bf 74}, 5036 (1995).}

\bibitem{3} C.R.Menyuk, R.Schiek and L.Torner
            {\em Journ.Opt.Soc.Am.B {\bf 11}, 2434-2443 (1994).}

\bibitem{4} L.Torner, C.R.Menyuk, W.E.Torruellas and G.I. Stegeman, 
{\em Opt.Lett. {\bf 20}, 13
(1995).}

\bibitem{5} A.V.Buryak and Yu.S.Kivshar, {\em Opt.Lett.{\bf 19}, 1612-1614
(1994).}

\bibitem{6} M.J.Werner and P.D.Drummond, {\em Journ.Opt.Soc.Am.B {\bf
10},2390-2393 (1993).}

\bibitem{7} M.J.Werner and P.D.Drummond, {\em Opt.Lett. {\bf 19}, 613-615
(1994).}

\bibitem{8} P.Ferro and S.Trillo, {\em Phys.Rev.E {\bf 51}, 4994-4998 (1995).}

\bibitem{9} G.I.Stegeman, M.Sheik-Bhae, E.Van Strykland and G.Assanto,
{\em Opt.Lett. {\bf 18}, 13-15 (1993).}

\bibitem{10} A.V.Buryak and Yu.S.Kivshar, {\em Phys.Lett.A {\bf 197}
407-412 (1995).}

\bibitem{11} A.V.Buryak and Yu.S.Kivshar, {\em Phys.Rev.A {\bf 51} R41-R44
(1995).}

\bibitem{12} A.V.Buryak and Yu.S.Kivshar, {\em Opt.Lett.{\bf 20}, 834-836
(1995).}

\bibitem{13}{M.J.Ablowitz, A.Ramani and H.Segur, {\em Journ.Math.Phys. {\bf
21}, 715 (1980).}}

\bibitem{14} M.J.Ablowitz and H.Segur, {\em Solitons and the Inverse 
Scattering Transform,
SIAM, Philadelphia, (1981).}

\bibitem{15} P.J.Olver, {\em Applications of Lie Groups to Differential
Equations, Springer, Berlin (1986).}

\bibitem{16} B.Champagne, W.Hereman and P.Winternitz, {\em Comput.Phys.Commun. 
{\bf 66},
319 (1991).}

\bibitem{17} P.Winternitz, {\em Lie Groups and Solutions of Nonlinear Partial 
Differential Equations, in:
Integrable Systems, Quantum Groups and Quantum Field Theory, Kluwer, Dordrecht 
(1993).}

\bibitem{18} D.Rand and P.Winternitz, {\em Comput.Phys.Commun. {\bf 42}, 359 
(1986).}

\bibitem{19} P.F.Byrd and M.D.Friedmann, {\em Handbook of Elliptic Integrals 
for Engineers and
Scientists, Springer, New York (1971).}

\bibitem{20} D.Levi, C.R.Menyuk and P.Winternitz, {\em Phys.Rev.A {\bf 44}, 
6057 (1991);
{\bf 49}, 2844 (1994).}

\bibitem{21} C.R.Menyuk, D.Levi and P.Winternitz, {\em Phys.Rev.Lett. {\bf 69}, 
3048
(1992).}

\bibitem{22} D.Levi, C.R.Menyuk and P.Winternitz, (editors). {\em
Self-Similarity in Stimulated Raman Scattering (Publications CRM, Montr\'eal,
1994).}



\end{thebibliography}
\end{document}